\documentclass[conference]{IEEEtran}

\usepackage{graphicx}
\usepackage{amsmath}
\usepackage{cite}
\usepackage{soul,xcolor}
\usepackage{booktabs}
\usepackage{cite}
\usepackage{longtable}
\usepackage{xurl}
\usepackage{makecell}
\usepackage{color,soul} 
\usepackage{adjustbox}
\usepackage{threeparttable}
\usepackage{ragged2e}  
\begin{document}

\title{Hardware-Matched Spiking Neural Networks for Real-Time Purkinje Cell Spike Classification}
\author{
\IEEEauthorblockN{
\begin{tabular}{c}
Arooba Farhan\IEEEauthorrefmark{1},
Afsah Syed\IEEEauthorrefmark{1},
Muhammad Fahd Ibrahim Bhatti\IEEEauthorrefmark{1},
Muhammad Areeb Khalid Kidwai\IEEEauthorrefmark{1},\\
Muhammad Ahsan Salar Khan\IEEEauthorrefmark{1},
Vincenzo Romano\IEEEauthorrefmark{3},
Muhammad Ali Siddiqi\IEEEauthorrefmark{1}\IEEEauthorrefmark{3}
\end{tabular}
}
\IEEEauthorblockA{
\IEEEauthorrefmark{1}Electrical Engineering Department, Lahore University of Management Sciences, Pakistan\\
\IEEEauthorrefmark{3}Department of Neuroscience, Erasmus University Medical Center, Rotterdam, The Netherlands\\
}
}

\maketitle

\begin{abstract}
Distinguishing Purkinje cell (PC) simple spikes (SSs) from complex spikes (CSs) is key to studying cerebellar function. PC recordings commonly rely on wired headstages, whose tethers restrict natural movement in freely behaving animals. Untethered headstages remove this constraint, but cannot continuously transmit raw recordings within their power budgets, making on-headstage spike classification an attractive form of compression, provided it runs accurately in real time on constrained hardware. This work presents two compact spiking neural networks (SNNs) matched to their target hardware: ConvSNN, which maps to the Matrix Vector Processor (MVP) of an EFM32PG28 microcontroller, and StreamSNN, which processes one waveform value per timestep using learned time constants and fixed synaptic delays for a dedicated SKY130 circuit. Evaluated on cells unseen during training, ConvSNN achieves a CS F1 of 0.873 with 1,018 parameters and StreamSNN 0.837 with 527 parameters, both with eight-bit weights. On the microcontroller, MVP acceleration reduces ConvSNN latency by 62.3\% and extends estimated battery life to 33.73~hours on a 50~mAh battery. In SKY130, StreamSNN occupies 0.173~mm$^2$ and consumes 0.46~$\mu$J per event, with 0.2~ms of computation within an 8.3~ms end-to-end latency.
\end{abstract}

\begin{IEEEkeywords}
Spiking neural networks, Purkinje cells, spike sorting, neuromorphic computing, matrix vector processor
\end{IEEEkeywords}

\section{Introduction}

\begin{figure*}[!t]
    \centering
    \includegraphics[width=1\textwidth]{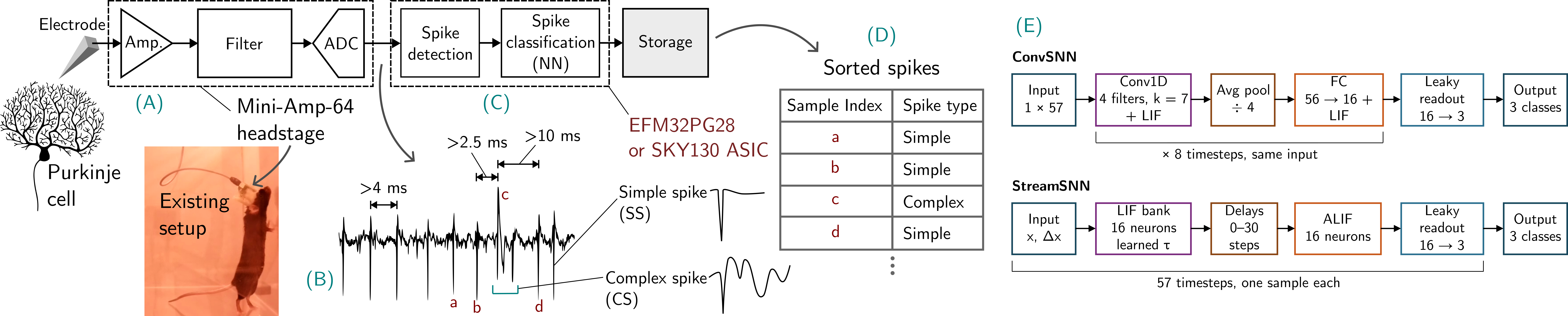}
    \caption{Overview of the proposed system. (A) Conventional wired setup, in which a mouse with a Mini-Amp-64 headstage records PC activity through an amplifier, filter, and ADC. (B) Extracellular PC recording containing simple spikes (SSs) and complex spikes (CSs). (C) On-device spike detection and classification. (D) Output stored as sample index and spike type instead of raw data. (E) The two SNN classifiers: ConvSNN processes the full 57-value window as a fixed input over eight timesteps, while StreamSNN processes one waveform value per timestep through LIF neurons with learned time constants, fixed synaptic delays, and adaptive-threshold (ALIF) neurons.}
    \label{fig:arch}
\end{figure*}

Purkinje cells (PCs), the sole output neurons of the cerebellar cortex, are central to motor coordination and learning~\cite{dezeeuw2011spatiotemporal}. Their extracellular activity contains simple spikes (SSs) and complex spikes (CSs), which arise from different synaptic inputs and carry distinct information about cerebellar processing~\cite{romano2020functional} (see Fig.~\ref{fig:arch}B). PC recordings are commonly acquired with wired headstages, whose tethers restrict natural movement in freely behaving animals~\cite{de2020ninscope} (see Fig.~\ref{fig:arch}A). Untethered headstages remove this constraint but face a data bottleneck: one channel sampled at 24.414~kHz with 16-bit resolution produces approximately 390~kb/s, making continuous raw-data transmission costly for a battery-powered device and requiring on-headstage compression within tight memory and power budgets.

Spike sorting, which detects spikes and assigns them to classes, offers an extreme form of such compression by reducing each event to a spike type and sample index (Fig.~\ref{fig:arch}C and D). However, many existing PC spike-sorting methods~\cite{markanday2020using,ZUR2019108423,sedaghat2021p,michikawa2021spike,owa2026spikesorting} are computationally demanding, require resources unavailable on constrained embedded platforms, or rely on both local field potentials (LFPs) and single-unit action potentials (APs). Recent works have demonstrated SS/CS classification on a mouse headstage ASIC~\cite{siddiqi2024lightweight} and on a microcontroller neural accelerator~\cite{fatima2025realtime}, establishing the feasibility of embedded spike sorting. However, these works do not evaluate generalization to unseen PCs, nor do they explore spiking neural networks (SNNs), whose sparse, event-driven computation may further reduce energy.

SNN neurons keep an internal state and communicate through discrete spikes~\cite{eshraghian2023training}, so computation for inputs that do not spike can be skipped. However, SNNs must also update each neuron's state at every timestep, and fewer spikes do not always mean lower hardware cost~\cite{bhattacharjee2024energy}. Their efficiency therefore depends on the network design, how often neurons spike, and how the computation maps to the target hardware.

In this work, we develop two compact SNNs that classify detected PC events as SSs, CSs, or false positives (FPs), i.e., detections that do not match an annotated spike. Both share a causal detection and preprocessing pipeline but are matched to different hardware targets. ConvSNN processes the full event window at once using shared convolutional filters and spiking neurons, producing regular matrix operations suited to the Matrix Vector Processor (MVP) of the EFM32PG28 microcontroller. StreamSNN instead processes the window one sample per timestep using spiking neurons with learned time constants and fixed synaptic delays, suited to a dedicated digital circuit in the SkyWater 130~nm (SKY130) process. We evaluate both SNNs on recordings from cells unseen during training. The main contributions of this work are:
\begin{enumerate}
\item \textbf{ConvSNN for microcontroller acceleration.} An SNN whose shared convolutional filters produce regular matrix operations suited to the EFM32PG28 MVP, achieving 0.873 CS F1 with 1,018 parameters.

\item \textbf{StreamSNN for dedicated hardware.} An SNN that processes one sample per timestep using leaky neurons with learned time constants and fixed synaptic delays, suited to compact digital logic, achieving 0.837 CS F1 with 527 parameters.

\item \textbf{Embedded implementation and evaluation.} We implement ConvSNN on the EFM32PG28 and StreamSNN in SKY130 with a low-cost causal noise-estimation front-end, report latency, energy, and area, and show that both SNNs keep their accuracy with eight-bit weights.
\end{enumerate}

\section{System Design and Implementation}

The proposed system detects spikes using a nonlinear energy operator (NEO)~\cite{mukhopadhyay1998new,yang2016hardware} on recordings normalized using only past samples. Each detection is aligned to its trough and defines a fixed waveform window, which is classified as SS, CS, or FP by ConvSNN or StreamSNN (Fig.~\ref{fig:arch}E).

\subsection{Recorded Data and Preprocessing}
\label{sec:exp-setup}

We use 64 manually annotated recordings from mouse PCs, each containing one cell, acquired at 24.414~kHz using a Mini-Amp-64 headstage, silicon probes, and an Intan RHD2000 amplifier~\cite{cambridge2020manual,intan2023manual}. Noise levels vary by more than 70$\times$ across recordings, from 10 to 775 ADC units, so the same spike can appear at very different amplitudes depending on the recording. We therefore normalize each recording using only past samples $x$. The baseline is estimated as the median, $m=\mathrm{median}(x)$, and the noise level as $\sigma=\mathrm{median}(|x-m|)/0.6745$, which converts the median absolute deviation to a standard deviation for Gaussian noise~\cite{quiroga2004unsupervised}. Medians are used because, unlike the mean and standard deviation, they are largely unaffected by spikes. The estimates are computed over a sliding 2~s window that is updated every 0.5~s, and each sample is normalized as $(x-m)/\sigma$.

\subsection{NEO Detection and Waveform Extraction}
\label{sec:detection}

Candidate events are detected using the nonlinear energy operator (NEO)~\cite{siddiqi2024lightweight,yang2016hardware,mukhopadhyay1998new}. NEO computes a sample-wise energy that grows with both signal amplitude and frequency, so it emphasizes sharp, high-amplitude spikes over slower background activity. After light smoothing of the normalized trace, an event is detected when this energy exceeds a fixed multiple of its recent average, followed by a short dead time to prevent repeated detections of the same spike.

\begin{figure}[!t]
    \centering
        \centering
        \includegraphics[width=1\linewidth]{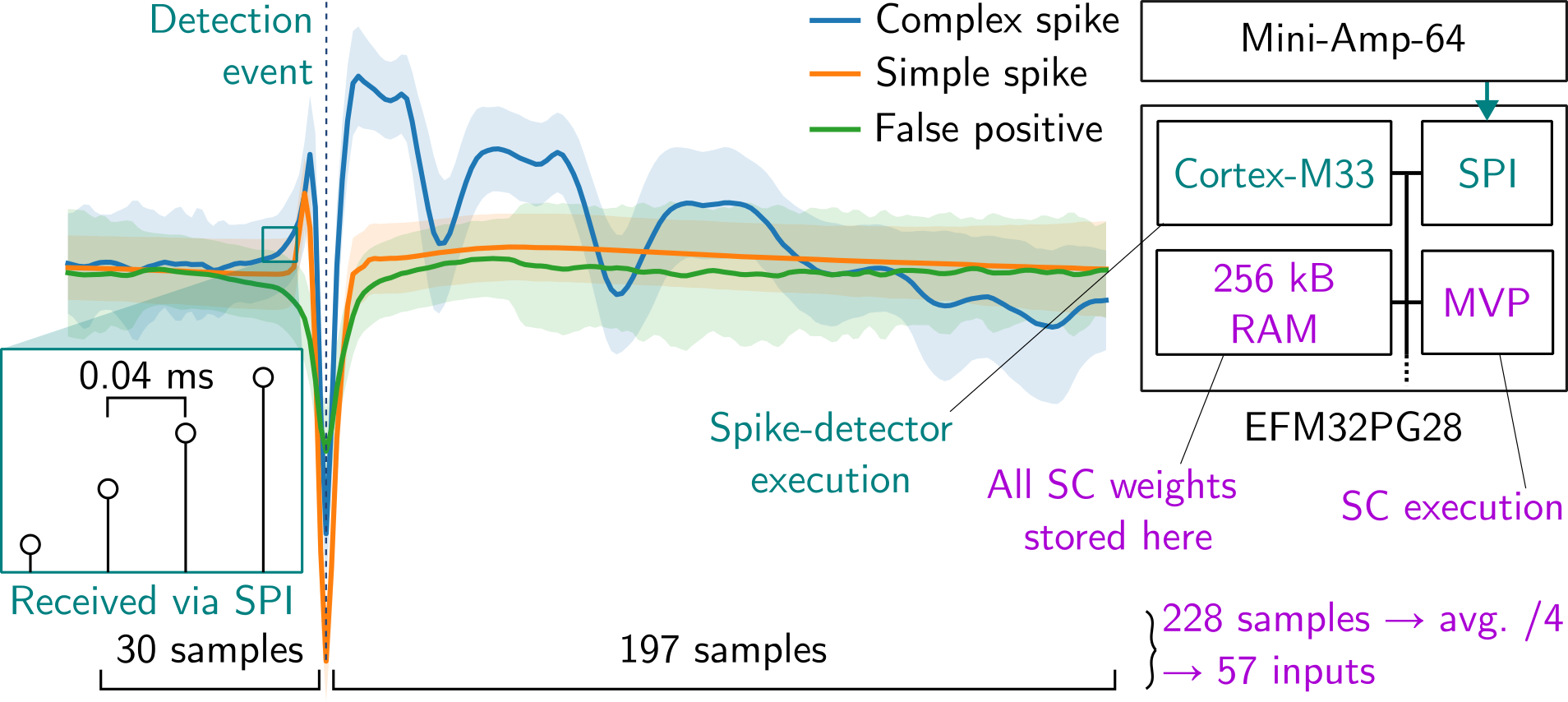}
        \caption{NEO-based spike detection and waveform extraction. Each detection defines a window spanning 30 samples before and 197 samples after the trough, reduced to 57 input values by four-sample averaging. Right: Functional mapping of ConvSNN onto the EFM32PG28 Cortex-M33 and MVP.}
        \label{fig:fig2}
\end{figure}

Each detection is aligned to the trough of the spike, i.e., its most negative sample. The waveform window spans 30 samples before and 197 samples after the trough ($\approx$9.4~ms) and is reduced to 57 values by four-sample averaging (see Fig.~\ref{fig:fig2}). Each detection is labeled by matching it to an annotated spike within $\pm0.5$~ms. If both an SS and a CS fall within this range, the CS label is used, and detections with no matching annotation are labeled FPs.

\subsection{SNN Architectures and Training}
\label{sec:snn}

\subsubsection{ConvSNN}
ConvSNN receives the 57-value waveform as a fixed input over eight timesteps. Four convolutional filters, each seven values wide, extract local features and drive leaky integrate-and-fire (LIF) neurons, which spike when their accumulated membrane potential crosses a threshold. Over the timesteps, each neuron converts its input strength into a spike count, and since the input is fixed, the convolution is computed once and reused. The spikes are averaged over groups of four positions to give 56 features, which feed 16 hidden LIF neurons and three non-spiking outputs; the class with the highest time-averaged output is selected. Since the hidden layers receive binary spikes, their synaptic operations reduce to additions that can be skipped for inactive inputs.

\subsubsection{StreamSNN}
Unlike ConvSNN, StreamSNN processes the 57 waveform values sequentially, one per timestep ($\approx$164~$\mu$s). Its 16 input LIF neurons have learned decay rates, capturing waveform features at different timescales. Each input neuron receives a weighted sum of the current and previous waveform values, $I_{i,t}=g_{i,0}z_t+g_{i,1}z_{t-1}$, where $g_{i,0}$ and $g_{i,1}$ are learned weights, so it can respond to both the level and the local slope of the waveform. Spikes from the input layer reach 16 hidden LIF neurons through fixed, randomly assigned delays of 0--30 timesteps (up to $\approx$4.9~ms), and the hidden neurons raise their firing threshold after each spike~\cite{hammouamri2024delays,yin2021adaptive}. Three non-spiking outputs use learned timestep-dependent weights, so the decision depends on when spikes occur as well as how many.

\subsubsection{Training and Evaluation}
Both networks are trained using backpropagation through time, with an arctangent surrogate gradient replacing the zero derivative of the spike threshold during the backward pass~\cite{eshraghian2023training}. Training uses AdamW, a OneCycle learning-rate schedule peaking at $3\times10^{-3}$, and a batch size of 512. To address class imbalance, the cross-entropy loss is weighted by $n_c^{-1/2}$, where $n_c$ is the number of training events in class $c$. ConvSNN and StreamSNN are trained for 20 and 16 epochs, respectively, and the checkpoint with the highest validation macro-F1 minus validation loss is kept. Weights are then quantized to 8 bits without retraining.

Our primary evaluation (cell-held-out) uses five-fold cross-validation over recordings: each fold tests on twelve or thirteen recordings (64 in total across the five folds), validates on eight, and trains on the remaining 43--44, so every cell is tested once and never used for training or model selection in its own fold. Since prior works use random event-level splits, where events from the same cell can appear in both training and test sets~\cite{owa2026spikesorting,fatima2025realtime}, we also report their 80/10/10 and 40/20/40 splits (random-split) for comparison only. Both models use identical splits. Per-class F1 is computed from precision and recall pooled over all detected test events, and macro-F1 is the mean of the SS, CS, and FP scores. Annotated spikes missed by the detector are excluded from these scores.

\subsection{Hardware Implementation}
\label{sec:impl-platform}

\subsubsection{Microcontroller}
ConvSNN is deployed on the Silicon Labs EFM32PG28, which combines an Arm Cortex-M33 with an MVP that computes in FP16. Eight-bit weights are stored in flash and expanded to FP16 in RAM at start-up. Average pooling is folded into the weights of the first fully connected layer, and the convolution is expressed as a dense matrix, so each inference uses 17 MVP calls: one for the convolution, whose output is reused across timesteps, and eight for each fully connected layer. The Cortex-M33 performs the LIF updates, skipping convolution neurons that cannot reach threshold, and accumulates the output over the eight timesteps. For comparison, the same network is also run on the Cortex-M33 alone. The deployed model is trained on a random 70/20/10 event-level split.

\subsubsection{ASIC}
StreamSNN is implemented in fixed point, with eight-bit weights and decay factors and 16-bit neuron states with eight fractional bits. The LIF neuron circuit is based on the design of Deng et al\cite{bitshift_LIF}. The input filters and neuron updates require multiplications, but each delayed spike from the input layer adds only a weight to the hidden neuron's membrane potential. The fixed delays are implemented by storing each input neuron's recent spikes in a shift register. Although StreamSNN processes samples sequentially, each classification is triggered by a NEO detection, so samples preceding the trough must be buffered. A front end estimates the running median and noise scale over the last 2~s for input normalization from logarithmically spaced block histograms, which need far fewer bits than storing the samples themselves, while a ring buffer retains recent offset-corrected samples so that each event window is available once the trough is detected. Overlapping events are queued (eight entries of about 64~bits) and processed one at a time by a single network context, whose approximately 175~B of state, excluding the 687~B of model parameters, comprises mainly neuron potentials and adaptation values (96~B) and the delay shift registers (62~B).

\section{Results}

\begin{table}[!t]
\centering
\caption{Classification F1 of proposed and prior classifiers.}
\label{tab:classification}
\scriptsize
\setlength{\tabcolsep}{2pt}
\begin{tabular}{@{}llrrrrr@{}}
\toprule
Work & Model & Params & Size & SS & CS & FP \\
\midrule
\multicolumn{7}{@{}l}{\textit{Cell-held-out}$^{a}$} \\
Ours & ConvSNN & 1,018 & 1,041 B & .958 & .873 & .911 \\
Ours & StreamSNN & 527 & 687 B & .950 & .837 & .894 \\
\midrule
\multicolumn{7}{@{}l}{\textit{Random split, 40/20/40}} \\
\cite{fatima2025realtime} & SNet-4 int8 & 2,500 & 2.5 kB & .965 & .867 & -- \\
\cite{fatima2025realtime} & SNet-4 float & 2,500 & -- & .970 & .890 & -- \\
Ours & ConvSNN & 1,018 & 1,041 B & .967 & .906 & .930 \\
Ours & StreamSNN & 527 & 687 B & .956 & .876 & .907 \\
\midrule
\multicolumn{7}{@{}l}{\textit{Random split, 80/10/10}} \\
\cite{owa2026spikesorting} & ResNet-18 & -- & -- & .975 & .892 & -- \\
Ours & ConvSNN & 1,018 & 1,041 B & .967 & .910 & .930 \\
Ours & StreamSNN & 527 & 687 B & .958 & .874 & .912 \\
\bottomrule
\end{tabular}\\[2pt]
\parbox{\columnwidth}{\scriptsize $^{a}$Five-fold, pooled over all held-out recordings. Our results use eight-bit weights. Dashes indicate unreported results.}
\end{table}

\begin{figure}[!t]
    \centering
    \includegraphics[width=\columnwidth]{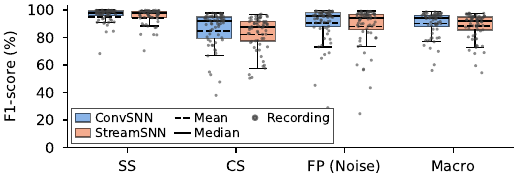}
    \caption{Per-recording F1 scores of ConvSNN and StreamSNN across 64 PCs under cell-held-out evaluation.}
    \label{fig:f1a}
\end{figure}

\subsection{SNN Classification Performance}
Under cell-held-out evaluation, ConvSNN achieves a CS F1 of 0.873 over all held-out events, compared with 0.837 for StreamSNN (Table~\ref{tab:classification}), so StreamSNN trades 0.036 CS F1 for 48\% fewer parameters. Computed per recording, the median CS F1 is 0.918 for ConvSNN and 0.876 for StreamSNN, while the lowest scores are 0.037 and 0.046, respectively, on the same recording, which contains only 25 CS events (Fig.~\ref{fig:f1a}).

Table~\ref{tab:classification} also compares our random-split results with published PC classifiers that use the same detector and split ratios. Under the 40/20/40 split, ConvSNN achieves a CS F1 of 0.906 (mean of three seeds), compared with 0.867 for the deployed SNet-4~\cite{fatima2025realtime}. Under 80/10/10, ConvSNN and StreamSNN achieve 0.910 and 0.874, compared with 0.892 for ResNet-18~\cite{owa2026spikesorting}.

Eight-bit post-training quantization has little effect under cell-held-out evaluation: CS F1 changes from 0.8737 to 0.8729 for ConvSNN and from 0.8384 to 0.8371 for StreamSNN, and float and int8 decisions agree on 99.31\% and 98.59\% of events, respectively.

\begin{table}[!t]
\centering
\caption{Spike-Classification Performance With/Without MVP}
\label{tab:spike-classification-mvp}
\scriptsize
\setlength{\tabcolsep}{3pt}
\begin{tabular}{@{}lrrrrrr@{}}
\toprule
\begin{tabular}[c]{@{}c@{}}Freq.\\(MHz)\end{tabular}
& \multicolumn{3}{c}{With MVP}
& \multicolumn{3}{c}{Without MVP} \\
\cmidrule(lr){2-4}\cmidrule(lr){5-7}
& $I_{\mathrm{avg}}$ (mA) & $t_{\mathrm{inf}}$ (ms) & $E_1$ ($\mu$J)
& $I_{\mathrm{avg}}$ (mA) & $t_{\mathrm{inf}}$ (ms) & $E_1$ ($\mu$J) \\
\midrule
13 & 0.83 & 5.620 & 15.440 & 0.85 & 14.557 & 40.957 \\
16 & 0.92 & 4.566 & 13.904 & 0.95 & 11.828 & 37.194 \\
19 & 1.01 & 3.845 & 12.854 & 1.04 & 9.9603 & 34.287 \\
26 & 1.20 & 2.810 & 11.257 & 1.24 & 7.279 & 29.876 \\
32 & 1.37 & 2.283 & 10.456 & 1.42 & 5.914 & 27.797 \\
80 & 2.94 & 0.9286 & 9.009 & 3.07 & 2.465 & 24.973 \\
\bottomrule
\end{tabular}

\parbox{\columnwidth}{\scriptsize
$I_{\mathrm{avg}}$: Avg. current consumption;
$t_{\mathrm{inf}}$: Inference time;
$E_1$: Energy cons. per inference.
}
\end{table}

\subsection{MCU Implementation}

\textit{Real-time performance:} Table~\ref{tab:spike-classification-mvp} compares classification latency on the Cortex-M33 alone with MVP-assisted execution. Both configurations operate at the lowest-current clock frequency that meets the 2.5~ms inference deadline, determined by the minimum interspike interval (Fig.~\ref{fig:arch}B): 32~MHz for the MVP-assisted implementation and 80~MHz for the CPU-only implementation. At equal clock (80~MHz), the MVP reduces classifier latency from 2.465~ms to 0.929~ms (62.3\%), allowing the MVP-assisted version to meet the deadline at 32~MHz (2.283~ms), while NEO detection takes approximately 9~$\mu$s at 80~MHz and 23~$\mu$s at 32~MHz. Although spike-driven execution would need about 3,400 synaptic operations per event instead of 9,100 dense multiply-accumulates, the MVP executes the layers densely, so the speedup comes from acceleration rather than spike sparsity. Acquiring the post-trough waveform takes approximately 8.1~ms, in addition to computation time. On the device, the int8 weights occupy 16.4~kB of flash, larger than the 1,041~B model because the convolution is stored as a dense matrix, and are expanded to a 32.8~kB FP16 copy in RAM for the MVP.

\begin{figure}[!t]
    \centering
    \includegraphics[width=0.7\columnwidth]{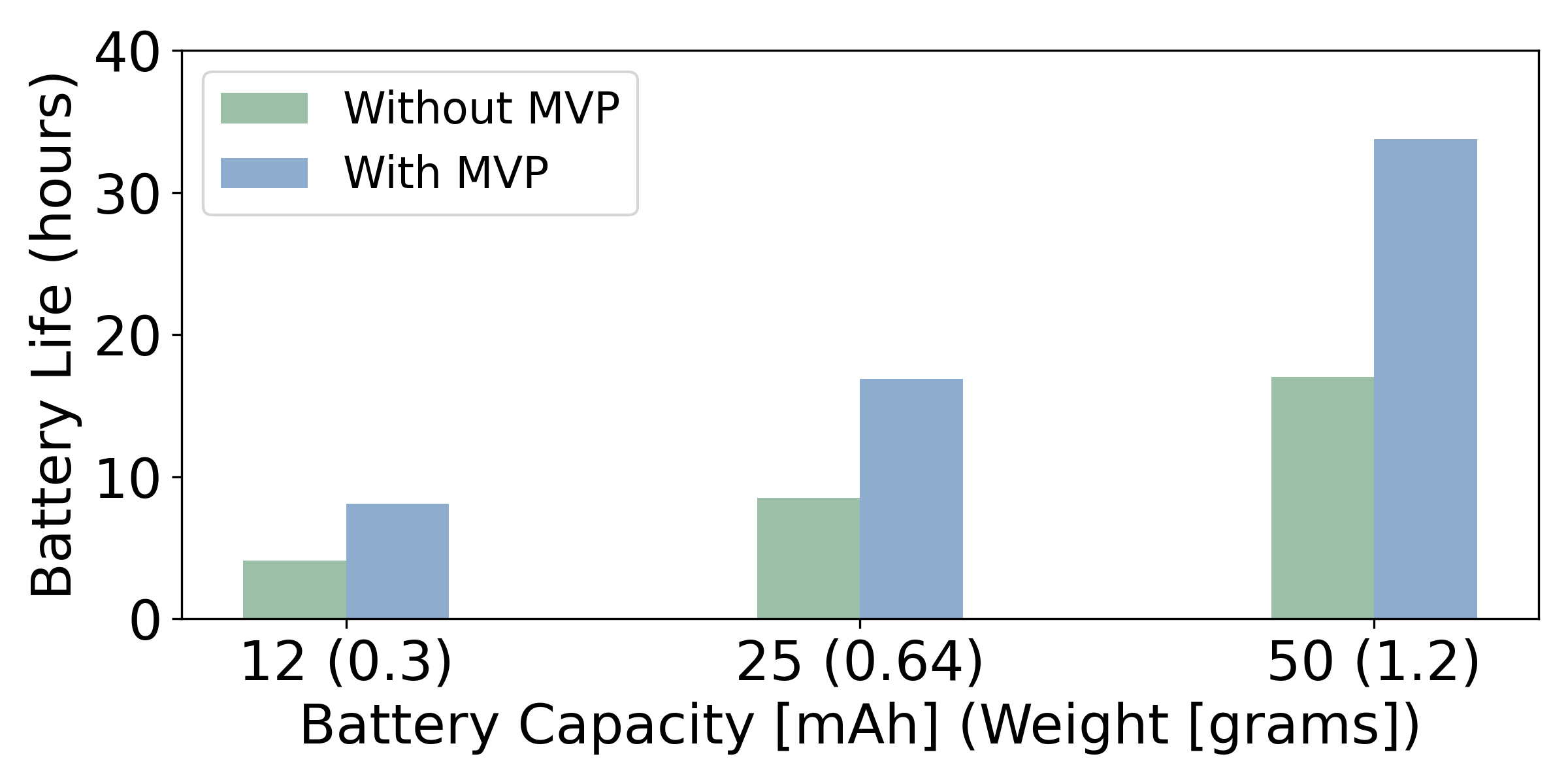}
    \caption{Estimated battery life for the complete head-stage using battery capacities of 12, 25, and 50 mAh.}
    \label{fig:battery_life_32MHz}
\end{figure}

\textit{Device autonomy:} Battery life was estimated from current measured with the Energy Profiler, including the classification and detector workloads and the average current of the ADC and STT-RAM~\cite{Everspin} (Fig.~\ref{fig:battery_life_32MHz}). By meeting the deadline at a lower clock, the MVP lowers the average current. For a 50~mAh battery, the estimated operating time is 17.04~hours for the CPU-only version and 33.73~hours with the MVP. This shows that MVP assisted implementation provides nearly twice the battery life of the CPU-only version.

\subsection{ASIC Implementation}
\label{sec:asic-results}

Per event, StreamSNN requires approximately 5,800 small integer multiplications for input filtering and neuron updates, while its hidden synapses perform only additions, approximately 5,300 spike-driven operations compared with 19,300 dense multiply-accumulates. On the 80/10/10 random split, the bit-exact fixed-point reference of the ASIC datapath achieves SS/CS/FP F1 of 0.959/0.872/0.911, within 0.002 of the eight-bit model in Table~\ref{tab:classification}, and agrees with floating-point decisions on 98.76\% of 160,274 test events. On the same events, the histogram-based noise estimator preserves 99.74\% of decisions relative to the exact median. Event windows overlap by 1.9 on average, with 99.9\% of events overlapping at most six windows and at most eight observed.

Table~\ref{tab:asic} reports post-layout results from LibreLane in SKY130 at a 100~ns clock period. Network energy per event is power multiplied by 1,995 clock cycles, and detector energy per sample uses one 24.414~kHz sample period. The latency of 8.3~ms is dominated by the 8.1~ms wait for the 197 post-trough samples. Network computation adds 0.2~ms, and the four-stage detector pipeline (0.16~ms) overlaps with this wait.

\begin{table}[!t]
\centering
\caption{ASIC results in SKY130 at a 100~ns clock period.}
\label{tab:asic}
\scriptsize
\setlength{\tabcolsep}{3pt}
\begin{tabular}{@{}lrrr@{}}
\toprule
Metric & Classifier & Detector & Whole chip \\
\midrule
Area (mm$^2$) & 0.173 & 0.047 & 1.59 \\
Power (mW) & 2.30 & 1.49 & 15.95 \\
Latency (ms) & 0.20 & 0.16 & 8.31 \\
Energy per event ($\mu$J) & 0.46 & -- & 3.20 \\
Energy per sample (nJ) & -- & 61 & -- \\
\bottomrule
\end{tabular}
\end{table}

\section{Conclusion}
This work presented two hardware-matched SNNs for real-time PC spike classification. Evaluated on cells unseen during training, ConvSNN achieves a CS F1 of 0.873 with 1,018 parameters and StreamSNN 0.837 with 527 parameters, both with eight-bit weights, and under the random-split conventions of prior work ConvSNN improves CS F1 from 0.867 to 0.906 with less than half the model storage. On the EFM32PG28, MVP acceleration reduces ConvSNN latency by 62.3\% and extends estimated battery life from 17.04 hours to 33.73 hours, while in SKY130, StreamSNN occupies 0.173~mm$^2$ and consumes 0.46~$\mu$J per event, with latency dominated by waveform acquisition. These results show that SNN efficiency depends on the platform: the MVP gains from dense acceleration rather than spike sparsity, whereas the ASIC exploits binary spikes to replace hidden-layer multiplications with additions. Future work will validate the system in freely behaving animals.

\section*{Acknowledgment}
The authors used Claude (Anthropic) for grammar and text refinement; all technical content, results, and claims were produced and verified by the authors.

\bibliographystyle{IEEEtran}
\bibliography{references}

@article{sedaghat2021p,
  title={P-sort: an open-source software for cerebellar neurophysiology},
  author={Sedaghat-Nejad, Ehsan and Fakharian, Mohammad Amin and Pi, Jay and Hage, Paul and Kojima, Yoshiko and Soetedjo, Robi and Ohmae, Shogo and Medina, Javier F and Shadmehr, Reza},
  journal={Journal of neurophysiology},
  volume={126},
  number={4},
  pages={1055--1075},
  year={2021},
  publisher={American Physiological Society Rockville, MD}
}

@article{markanday2020using,
  title={Using deep neural networks to detect complex spikes of cerebellar Purkinje cells},
  author={Markanday, Akshay and Bellet, Joachim and Bellet, Marie E and Inoue, Junya and Hafed, Ziad M and Thier, Peter},
  journal={Journal of neurophysiology},
  volume={123},
  number={6},
  pages={2217--2234},
  year={2020},
  publisher={American Physiological Society Bethesda, MD}
}

@article{ZUR2019108423,
title = {Using extracellular low frequency signals to improve the spike sorting of cerebellar complex spikes},
journal = {Journal of Neuroscience Methods},
volume = {328},
pages = {108423},
year = {2019},
issn = {0165-0270},
doi = {https://doi.org/10.1016/j.jneumeth.2019.108423},
url = {https://www.sciencedirect.com/science/article/pii/S0165027019302808},
author = {Gil Zur and Mati Joshua}
}

@article{romano2020functional,
  title={Functional convergence of autonomic and sensorimotor processing in the lateral cerebellum},
  author={Romano, Vincenzo and Reddington, Aoibhinn L and Cazzanelli, Silvia and Mazza, Roberta and Ma, Yang and Strydis, Christos and Negrello, Mario and Bosman, Laurens WJ and De Zeeuw, Chris I},
  journal={Cell reports},
  volume={32},
  number={1},
  pages={107867},
  year={2020},
  publisher={Elsevier}
}

@inproceedings{siddiqi2024lightweight,
  title={A lightweight architecture for real-time neuronal-spike classification},
  author={Siddiqi, Muhammad Ali and Vrijenhoek, David and Landsmeer, Lennart PL and van der Kleij, Job and Gebregiorgis, Anteneh and Romano, Vincenzo and Bishnoi, Rajendra and Hamdioui, Said and Strydis, Christos},
  booktitle={Proceedings of the 21st ACM International Conference on Computing Frontiers},
  pages={32--40},
  year={2024}
}

@article{michikawa2021spike,
  title={A spike-sorting method based on hierarchical clustering to discriminate extracellularly recorded simple spikes and complex spikes from cerebellar Purkinje cells},
  author={Michikawa, Takayuki and Isobe, Keisuke and Itohara, Shigeyoshi},
  journal={bioRxiv},
  pages={2021--09},
  year={2021},
  publisher={Cold Spring Harbor Laboratory}
}

@manual{cambridge2020manual,
	title="{Mini-Amp-64 User Guide -- Version 1.0}",
	author="{Cambridge NeuroTech}",
	year={2020}
}

@article{yang2016hardware,
  title={Hardware efficient automatic thresholding for NEO-based neural spike detection},
  author={Yang, Yuning and Mason, Andrew J},
  journal={IEEE Transactions on Biomedical Engineering},
  volume={64},
  number={4},
  pages={826--833},
  year={2016},
  publisher={IEEE}
}

@manual{intan2023manual,
	title="{RHD2000 Series Digital Electrophysiology Interface Chips}",
	author="{Intan Technologies, LLC}",
	year={2023}
}

@article{mukhopadhyay1998new,
  title={A new interpretation of nonlinear energy operator and its efficacy in spike detection},
  author={Mukhopadhyay, Sudipta and Ray, GC},
  journal={IEEE Transactions on biomedical engineering},
  volume={45},
  number={2},
  pages={180--187},
  year={1998},
  publisher={IEEE}
}

@online{Everspin,
  author = "{Everspin Technologies}",
  title = "{Spin-transfer Torque MRAM Technology}",
  url = {https://www.everspin.com/},
  year={2024}
}

@inproceedings{owa2026spikesorting,
  author    = {Owais, Muhammad Faizan and Saleem, Hamza and Landsmeer, Lennart P. L. and Romano, Vincenzo and Strydis, Christos and Siddiqi, Muhammad Ali},
  title     = {Spike Sorting of Purkinje Cells via Adaptive Histogram Thresholding and Deep Learning},
  booktitle = {2026 IEEE Engineering in Medicine and Biology Conference (EMBC)},
  year      = {2026},
  organization = {IEEE}
}

@inproceedings{fatima2025realtime,
  author    = {Fatima, Kiran and Romano, Vincenzo and Strydis, Christos and Siddiqi, Muhammad Ali},
  title     = {Real-Time, Low-Power Purkinje Cell Spike Sorting for Headstage Systems in Freely Moving Mice},
  booktitle = {Proceedings of the 2025 IEEE Biomedical Circuits and Systems Conference (BioCAS)},
  pages     = {274--278},
  year      = {2025},
  organization = {IEEE},
  doi       = {10.1109/BioCAS67066.2025.00067}
}

@INPROCEEDINGS{bitshift_LIF,
  author={Deng, Qiaosong and Ma, Junyu and Cai, Hanfeng and You, Hao and Kanchwala, Mustafa and Xu, Jianxiong and Amirsoleimani, Amirali and Zariffa, José and Genov, Roman},
  booktitle={2024 IEEE Biomedical Circuits and Systems Conference (BioCAS)}, 
  title={Multiplierless Spiking Neural Network for Motor Signal Decoding in the Peripheral Nervous System}, 
  year={2024},
  volume={},
  number={},
  pages={1-5},
  doi={10.1109/BioCAS61083.2024.10798280}}

@inproceedings{bhattacharjee2024energy,
  author    = {Bhattacharjee, Abhiroop and Yin, Ruokai and Moitra, Abhishek and Panda, Priyadarshini},
  title     = {Are SNNs Truly Energy-Efficient? A Hardware Perspective},
  booktitle = {2024 IEEE International Conference on Acoustics, Speech and Signal Processing (ICASSP)},
  pages     = {13311--13315},
  year      = {2024},
  doi       = {10.1109/ICASSP48485.2024.10448269}
}

@article{eshraghian2023training,
  author  = {Eshraghian, Jason K. and Ward, Max and Neftci, Emre O. and Wang, Xinxin and Lenz, Gregor and Dwivedi, Girish and Bennamoun, Mohammed and Jeong, Doo Seok and Lu, Wei D.},
  title   = {Training Spiking Neural Networks Using Lessons From Deep Learning},
  journal = {Proceedings of the IEEE},
  volume  = {111},
  number  = {9},
  pages   = {1016--1054},
  year    = {2023},
  doi     = {10.1109/JPROC.2023.3308088}
}

@article{dezeeuw2011spatiotemporal,
  author  = {De Zeeuw, Chris I. and Hoebeek, Freek E. and Bosman, Laurens W. J. and Schonewille, Martijn and Witter, Laurens and Koekkoek, Sebastiaan K.},
  title   = {Spatiotemporal Firing Patterns in the Cerebellum},
  journal = {Nature Reviews Neuroscience},
  volume  = {12},
  number  = {6},
  pages   = {327--344},
  year    = {2011},
  doi     = {10.1038/nrn3011}
}

@article{yin2021adaptive,
  author = {Yin, Bojian and Corradi, Federico and Boht{\'e}, Sander M.},
  title = {Accurate and Efficient Time-Domain Classification with Adaptive Spiking Recurrent Neural Networks},
  journal = {Nature Machine Intelligence},
  volume = {3},
  number = {10},
  pages = {905--913},
  year = {2021},
  doi = {10.1038/s42256-021-00397-w}
}

@inproceedings{hammouamri2024delays,
  author = {Hammouamri, Ilyass and Khalfaoui Hassani, Ismail and Masquelier, Timoth{\'e}e},
  title = {Learning Delays in Spiking Neural Networks Using Dilated Convolutions with Learnable Spacings},
  booktitle = {International Conference on Learning Representations},
  year = {2024},
  url = {https://proceedings.iclr.cc/paper_files/paper/2024/hash/4df1cc5a7528b7197ad8ae76ff30107a-Abstract-Conference.html}
}

@article{de2020ninscope,
  title={NINscope, a versatile miniscope for multi-region circuit investigations},
  author={de Groot, Andres and van den Boom, Bastijn JG and van Genderen, Romano M and Coppens, Joris and van Veldhuijzen, John and Bos, Joop and Hoedemaker, Hugo and Negrello, Mario and Willuhn, Ingo and De Zeeuw, Chris I and others},
  journal={Elife},
  volume={9},
  pages={e49987},
  year={2020},
  publisher={eLife Sciences Publications, Ltd}
}

@article{quiroga2004unsupervised,
  title={Unsupervised spike detection and sorting with wavelets and superparamagnetic clustering},
  author={Quiroga, R Quian and Nadasdy, Zoltan and Ben-Shaul, Yoram},
  journal={Neural computation},
  volume={16},
  number={8},
  pages={1661--1687},
  year={2004},
  publisher={MIT Press}
}

\end{document}